\documentclass[
 reprint,
 amsmath,amssymb,
 aps,
]{revtex4-2}
\usepackage[T1]{fontenc}
\usepackage{graphicx}
\usepackage{dcolumn}
\usepackage{bm}
\usepackage{amsmath}
\usepackage{lmodern}
\usepackage{mathtools}
\usepackage{caption}
\usepackage{siunitx}
\usepackage{gensymb}
\usepackage{caption}
\usepackage{subcaption}
\usepackage{lipsum}
\usepackage{float} %required for the placement specifier H
\usepackage[colorlinks,citecolor=black,linkcolor=black,urlcolor=black]{hyperref}
\usepackage[normalem]{ulem}

\def\MoTe2{MoTe$_{\tiny{\textrm{2}}}$}
\def\SiO2{SiO$_{\tiny{\textrm{2}}}$}
\def\Al2O3{Al$_{\tiny{\textrm{2}}}$O$_{\tiny{\textrm{3}}}$}
\def\AlOx{AlO$_{\tiny{\textrm{x}}}$}
\def\Mo6Te6{Mo$_{\tiny{\textrm{6}}}$Te$_{\tiny{\textrm{6}}}$}
\def\MoSe2{MoSe$_{\tiny{\textrm{2}}}$}
\def\MoS2{MoS$_{\tiny{\textrm{2}}}$}
\def\WSe2{WSe$_{\tiny{\textrm{2}}}$}

\DeclareUnicodeCharacter{0327}{\textendash}
\DeclareUnicodeCharacter{0229}{\textendash}
\begin{document}

\title{Molecular Beam Epitaxy growth of MoTe$_{2}$ on Hexagonal Boron Nitride}
\author{B.~Seredyński$^1$}
\email{Bartlomiej.Seredynski@fuw.edu.pl}
\author{R.~Bożek$^1$}
\author{J.~Suffczyński$^1$}
\author{J. Piwowar $^2$}
\author{J.~Sadowski$^{1,3,4}$}
\author{W.~Pacuski$^1$}
\email{Wojciech.Pacuski@fuw.edu.pl}

\affiliation{$^1$Institute of Experimental Physics, Faculty of Physics, \\University of Warsaw, ul. Pasteura 5, PL-02-093 Warsaw, Poland}
\affiliation{$^2$ Biological and Chemical Research Centre, Faculty of Chemistry, University of Warsaw, ul. Żwirki i Wigury 101, PL 02-089 Warsaw, Poland }
\affiliation{$^3$The Institute of Physics, Polish Academy of Sciences,\\ al. Lotnik\'ow 32/46, PL-02-668 Warsaw, Poland}
\affiliation{$^4$Department of Physics and Electrical Engineering, Linnaeus University, SE-391 82
Kalmar, Sweden}

%%%%%%%%%%%%%%%%%%%%%%%%%%%%%%%%%%%%%%%%%%%%%%%%%%%%%%%%%%%%%%%%%%%%%%%%%%%%%%%%%%%%%%%%%%%%%%%%%

\begin{abstract}
    Hexagonal boron nitride has already been proven to serve as a decent substrate for high quality epitaxial growth of several 2D materials, such as graphene, \MoSe2, \MoS2 or \WSe2. Here, we present for the first time the molecular beam epitaxy growth of \MoTe2 on atomically smooth hexagonal boron nitride (hBN) substrate.
    Occurrence of \MoTe2 in various crystalline phases such as distorted octahedral 1T’ phase with semimetal properties or hexagonal 2H phase with  semiconducting properties opens a possibility of realisation of crystal-phase homostructures with tunable properties.
    Atomic force microscopy studies of \MoTe2 grown in a single monolayer regime enable us to  observe impact of growth conditions on formation of various structures: flat grains, net of one-dimensional structures, quasi continuous monolayers with bilayer contribution. Comparison of the distribution of the thickness with Poisson distribution shows that tested growth conditions favorites formation of grains with monolayer thickness.  The diffusion constant of \MoTe2 grown on hBN can  reaches order of ~10$^{-6}$~cm$^{2}$/s for typical growth conditions. Raman spectroscopy results suggest a coexistence of various phases with domination of 2H \MoTe2 for samples grown at lower temperatures. XPS measurements confirms the stoichiometry of \MoTe2.
    
    Keywords: A1.Diffusion; A1.Layers distribution; A1.Surface morphology; A3.Molecular beam epitaxy; B1.Molybdenum Ditelluride; B2.Transition Metal Dichalcogenides

\end{abstract}
\maketitle
%%%%%%%%%%%%%%%%%%%%%%%%%%%%%%%%%%%%%%%%%%%%%%%%%%%%%%%%%%%%%%%%%%%%%%%%%%%%%%%%%%%%%%%%%%%%%%%%%

\section{\label{sec:intro}Introduction} 
    Molybdenum ditelluride belongs to the Transition Metals Dichalcogenides (TMDC) family. It occurs in several thermodynamically stable crystalline phases, such as hexagonal (semiconducting) 2H, octahedral (metallic) 1T', and distorted octahedral (semimetalic) Td \cite{MoTe2_phase_stability} one. One should also note the existence of an interesting \Mo6Te6 \cite{Mo6Te6} phase that is not yet much investigated.
    The 1T' phase is exciting due to the theoretically predicted and experimentally confirmed topological semimetal (i.e., type-II Weyl semimetal)\cite{MoTe2_Weyl} properties, including the occurrence of Fermi arc surface states \cite{MoTe2_arc_1}, the giant magnetoresistance \cite{MoTe2_GMR} or the planar Hall effect \cite{MoTe2_Hall}. 
    From the other hand, the 2H-\MoTe2 phase is in particular interesting in a 1 ML thickness regime because of a direct energy gap, which value is close to 1.1~eV \cite{MoTe2_band_gap}. This allows to use semiconducting \MoTe2 in  optoelectronic devices. Other unique features of this phase are, e.g., hidden spin-polarized bands \cite{MoTe2_hidden_bands} or the highly tunable non-linear Hall effect induced by spin-orbit coupling \cite{MoTe2_Hall_przestrajalny}. 
    Due to the low bond energy between molybdenum and tellurium atoms \cite{MoTe2_ene_wiazania}, it is possible to force the \MoTe2 phase transition by external factors. Those can be stress introduced by external pressure \cite{MoTe2_fazy_naprezenie}, laser irradiation \cite{MoTe2_homojunctions_1}, or electrical gating \cite{MoTe2_fazy_electrostatic_doping}. Consequently, one of the most promising directions of \MoTe2 research is the creation of homogeneous ohmic contacts based on the semiconducting (2H) and metallic (1T') \MoTe2 phases \cite{MoTe2_homojunctions_1}.  
    All those attractive properties have been investigated so far mostly for exfoliated samples. This encourages developing epitaxial growth methods of \MoTe2. In the literature there are already some reports concerning \MoTe2 layers grown by Molecular Beam Epitaxy (MBE). 
    \MoTe2 was grown on two-dimensional materials such as \MoS2 \cite{MoTe2_MoS2}, or standard 3D substrate such as GaAs \cite{Furdyna_2, MoTe2_nasze} yielding either a mix of \MoTe2 phases or a pure 2H phase.  The studies of \MoTe2 grown on graphene \cite{ MoTe2_graphene_Chiny, MoTe2_BLG, MoTe2_druty_plasko} showed that the temperature of growth or annealing has an impact on the size of the areas of a specific phase. The growth of exactly 1 ML thick, polycrystalline \MoTe2 covering the entire surface of a 2-inch \Al2O3, or \SiO2/Si substrate was also achieved \cite{MoTe2_Al2O3, MoTe2_Dielectrics}. So far, the only pure 1T' phase MBE growth reported in the literature was performed on InAs substrate \cite{MoTe2_InAs}. However, these are still small monocrystalline islands.
    Possible direction of the development of \MoTe2 epilayers can be for example the growth on the hexagonal boron nitride (hBN) substrate, which is expected to offer many potential benefits for the epitaxy of TMDC. Ideally, the hBN plane can be considered as atomically smooth, that is, devoid of atomic steps. Then one can expect the absence of unsaturated bonds \cite{WSe2_hBN_1}. This should reduce the nucleation density, precluding island-type growth in several sites on the substrate. The expected growth mode should therefore be 2D layer-by-layer. Another argument in favor of using hBN as a substrate is its hexagonal crystal structure making it compatible with the epitaxial growth of most two-dimensional TMDC. Additionally, the diffusion of adatoms on the surface should increase. Last, but not least, the high dielectric constant of hBN \cite{MoS2_hBN} and its chemically inert surface suppresses scattering from charged contaminants and substrate phonons. This should result in a higher mobility and an increased exciton density in the grown TMDC layers \cite{MoSe2_nasze}.
In this work we present the results of growth with various growth parameters which are growth stage temperature, tellurium flux and annealing after the growth in order to find the best conditions for growth of flat and continuous \MoTe2 layer on hBN.
%%%%%%%%%%%%%%%%%%%%%%%%%%%%%%%%%%%%%%%%%%%%%%%%%%%%%%%%%%%%%%%%%%%%%%%%%%%%%%%%%%%%%%%%%%%%%

\section{\label{sec: Results} Results}
MBE growth of \MoTe2 layers on hBN was performed in a twin chamber MBE system (SVTA), one growth chamber is dedicated to II–VI, and the second one to III–V compounds. The \MoTe2 growth was performed in II-VI chamber.  Mo (99.995\% purity) has been evaporated from the electron beam source and Te (7N purity) from the dual-filament Knudsen cell.

\begin{figure*}
         \centering
         \includegraphics[width=\textwidth]{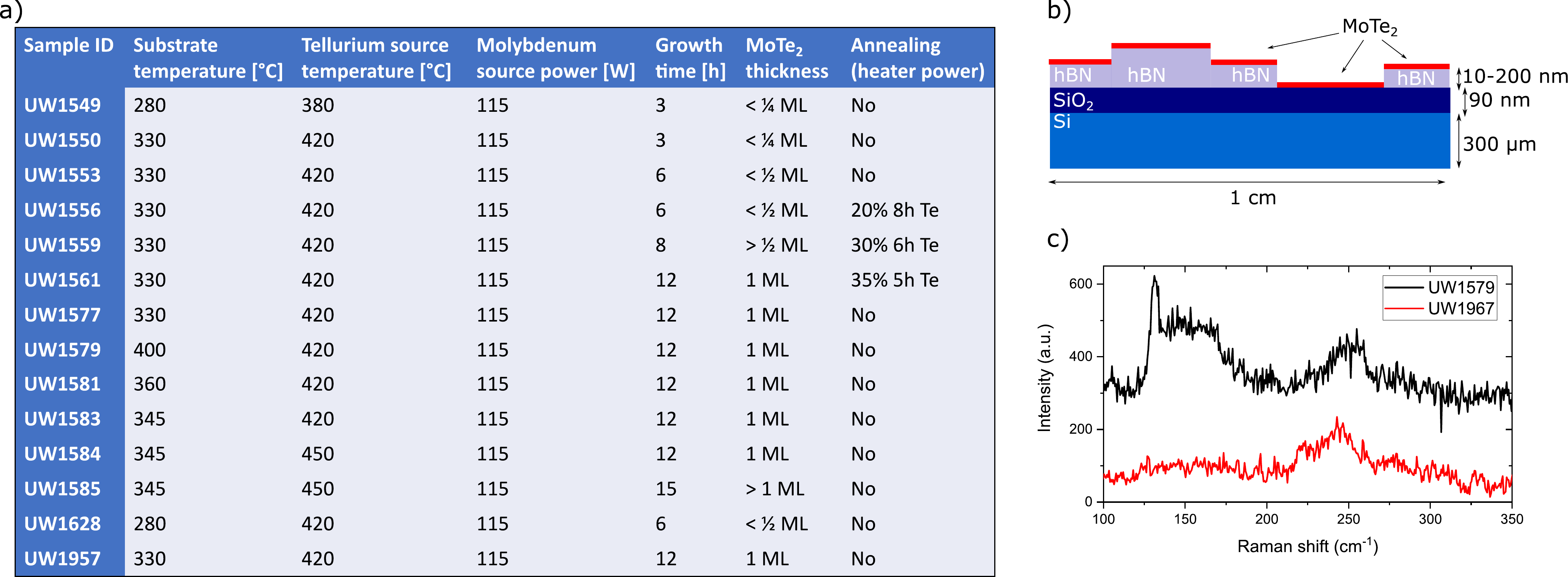}
    \caption{a) Table of samples studied in this work. Tellurium source temperature is the indication of the thermocouple at the tellurium source. A 20 degree increase in the tellurium source temperature corresponds to approximately two-fold increase of the tellurium flux. Molybdenum source power is the power of the supplier that controls the Mo e-beam source. The power of 115~W allows to grow about 1 monolayer per 12~h. The last column contains the information whether the sample has been annealed after the growth. The heater power defines the temperature of the substrate: 20\% corresponds to the indication of the thermocouple located on the substrate of about 400~\degree{}C, 30\% about 600~\degree{}C, 35\% about 700~\degree{}C. Annealing, if applied, takes place under the tellurium flux.Sample ID code is an abbreviation of University of Warsaw and the ordinary number of the sample.
    b) Scheme of the resulted \MoTe2 on hBN film. The \MoTe2 layers are formed also on the bare \SiO2/Si. The hBN flakes are of different thicknesses and lateral sizes.
    c)Example Raman spectra with multiple and spectrally wide lines that suggest occurrence of the mixture of various phases with domination of 2H-\MoTe2 for sample grown at lower substrate temperature (UW1967) and \Mo6Te6 for sample grown at high substrate temperature (UW1579).}
    \label{fig: tabela_i_schemat}
 \end{figure*}

    In order to grow layers on a monocrystal boron nitride, it was necessary first to prepare such a substrate. For this purpose the process of exfoliating hBN from bulk and depositing small (several dozen square micrometers) flakes on the \SiO2/Si surface was done. The \SiO2 substrates used by us are squares of 1 cm$^2$ size. Such surface comprises typically hundreds of hBN flakes. The flakes of hBN have different thicknesses and lateral sizes, which means that they do not uniformly and completely cover the entire \SiO2/Si surface.
    %After the substrate was placed on the holder and introduced into the vacuum chamber, \MoTe2 was grown. 
    Hence, while growing in the MBE chamber, Mo and Te molecules condense also on the uncovered areas of the \SiO2/Si surface.     The sample scheme is shown in Figure \ref{fig: tabela_i_schemat} b).
    
  After introduction of the substrate to MBE chamber, there was a degas procedure,  20 min at 700 \degree C, next substrate was cooled down to growth temperature in range 280-400 \degree C, which is typical for growth of telluride compounds. This range results from the fact that at lower temperatures Te condense, at higher temperatures, Te quickly desorbs. The table of the studied samples and their growth parameters such as e.g. substrate temperature is shown in Figure \ref{fig: tabela_i_schemat}. Growth was performed without interruptions, with both shutters of Mo and Te open. After growth selected samples has been annealed in Te flux, with Mo shutter closed. For all samples power of e-beam source with Mo has been constant, therefore amount of deposited Mo was directly proportional to the growth time, where 12h of deposition corresponds to about 1 ML of the  material.

    A convenient tool for determining the morphological properties of  \MoTe2 layers is atomic force microscopy (AFM). It is particularly effective in determining the coverage of samples grown in a thickness regime below a few atomic layers. One can then easily determine the coverage of the substrate surface, the distribution of successive layers with their thicknesses and the surface roughness. 
    The AFM images presented below cover the area of 2 $\mu$m $\times$2 $\mu$m and present the surface of \MoTe2 deposited entirely on a single hBN flake.

\begin{figure*}
         \centering
         \includegraphics[width=\textwidth]{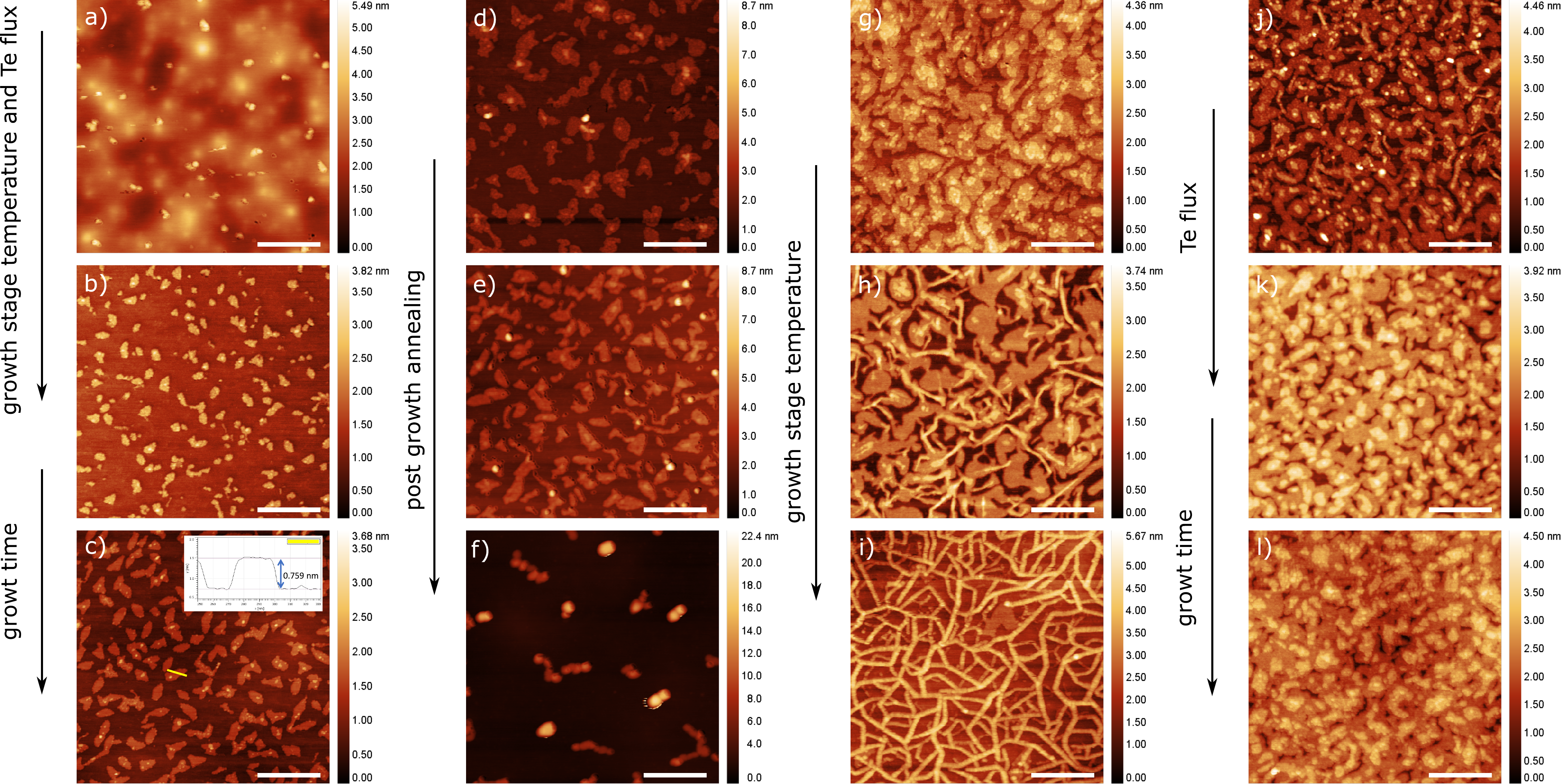}
    \caption{AFM images of \MoTe2 grown on hBN. Parameters varied for a given sample series are indicated next to the images.
    a) UW1549 - growth carried out at low temperature and with low Te flux. b) UW1550 - increase in the growth temperature and the flux of Te molecules. c) UW1553 - doubling the growth time results in the expansion of islands. Inset shows the thickness of the island corresponding to exactly one \MoTe2 monolayer. 
    d) - f) UW1556, UW1559, UW1561 - subsequent annealing at higher temperatures causes the edges of the \MoTe2 grains to curl, leading to the formation of convex and meandering structures.
    g) - i) UW1577, UW1581, UW1579 - higher growth temperature causes the formation of both flat island regions and one-dimensional structures, the highest of the proposed growth temperatures results in the formation of a mesh of one-dimensional structures, known as \Mo6Te6 \cite{Mo6Te6_referee}}
    j) and k) UW1583 and UW1584 - increasing the Te flux increases the smoothness of the layers. Figure l) UW1585 - extended growth time with a large Te flux. 
    The scale bar corresponds to 500~nm in the case of each figure.
    \label{fig: AFM_serie_pion}
 \end{figure*}

   Figure \ref{fig: AFM_serie_pion}a) shows the AFM image of the UW1549 sample grown at 280 \degree C and at relatively low Te flux. That is, the ratio of Mo to Te was about 1:17000. In this case small size of grains forming on the surface precludes reliable determination of their crystalline properties.  In order to check if amount of deposited Mo was too low or just nucleation of \MoTe2 has not yet started, we performed second growth keeping the same both Mo flux and the growth time (and consequently the same amount of deposited Mo), but the Te flux was increased 4 times and the growth temperature was increased to 330 \degree C. In such a case, much larger \MoTe2 grains were obtained (see UW1550 sample in Fig. \ref{fig: AFM_serie_pion}b))  indicating important improvement in effectiveness of nucleation. In this case the contrast between \MoTe2 and hBN is high enough to distinguish separate islands, but amount of \MoTe2 is still too low to approach covering of the whole surface, and consequently test of longer growth is needed.
    While maintaining the above growth conditions, but increasing the growth time twice, the expansion of grains was observed - the UW1553 sample in Fig. \ref{fig: AFM_serie_pion}c). The islands of \MoTe2 start to coalesce in some places. Analysis of the cross-section through the selected \MoTe2 grain (inset to the Fig. \ref{fig: AFM_serie_pion}c)) indicates that it is flat and its thickness corresponds to exactly one monolayer of \MoTe2. Some grains exhibit explicit white spots. These are the seeds of the second \MoTe2 monolayer. It indicates that the growth conditions are not yet optimal for obtaining a flat single monolayer.

    From our previous work on  MBE growth of \MoSe2 on hBN \cite{MoSe2_nasze, Ludwiczak} we concluded that annealing of the samples after the growth helps to achieve continuous and smooth area of the first TMDC monolayer. 
    The panels d)-f) of Fig. \ref{fig: AFM_serie_pion} show AFM scans of successive samples annealed at various temperatures after the growth. Figure \ref{fig: AFM_serie_pion}d) shows the UW1556 sample that was annealed about 70 degrees higher than the growth temperature. It can be seen that the individual grains have rounded edges and became more convex. The average size of the islands indeed increased with respect to the similar UW1553 sample, which was not annealed after the growth. Therefore, the next sample UW1559 shown in Fig. \ref{fig: AFM_serie_pion}e) was annealed 200 \degree C higher simultaneously prolonging the growth time by 50 \%. The edges of the grains are rounded. The size of the grains did not increase significantly. In the next sample, UW1561 shown in Fig. \ref{fig: AFM_serie_pion}f), the annealing temperature was increased by another 100 \degree C and the growth time increased by the factor of 1.5. The material forms meandering longitudinal islands far from being flat. A similar effect was reported in the literature for \MoTe2 grown on \MoS2 \cite{MoTe2_MoS2} or MoSe2 grown on hBN \cite{druty_MoSe2}.

    Growth temperature is another parameter investigated here. Intuitively higher temperature should enhance the migration of adatoms on the surface and promote the growth of a smooth layer. Thus, a series of growths at increasing  temperatures was performed. Simultaneously the growth time was brought closer to one monolayer which is 12 h for extremely low Mo flux used here. The growth time calibration was done in previous studies during \MoTe2 growths on GaAs substrate \cite{MoTe2_nasze}. For this series the samples were not annealed after the growth. 
    Successive AFM images of samples grown at various temperatures are shown in  Figures \ref{fig: AFM_serie_pion}g)-i). Figure  \ref{fig: AFM_serie_pion}g) shows the AFM scan of the surface of UW1577 sample. The growth conditions are identical as for the sample shown in Fig. \ref{fig: AFM_serie_pion}b), but the growth time was increased 4 times. The grains adhere to each other creating continuous areas of the large surface. Unfortunately, under these growth conditions, \MoTe2 tends to form the second and the third monolayers prior to the full coverage of the hBN substrate surface.
    Figure \ref{fig: AFM_serie_pion}h) shows the surface of the UW1581 sample. Here, the growth temperature has been raised by 30 \degree C with respect to the previous case. The other parameters were kept the same. The AFM scan shows areas with smooth surfaces interspersed with one-dimensional-like structures. The growth temperature increase did not lead to a more complete coverage of the first monolayer.
    With the increase of the growth temperature by another 40 \degree C for the  UW1579 sample, a network of one-dimensional objects was observed, as shown in Fig. \ref{fig: AFM_serie_pion}i). A kind of molybdenum telluride nanowires are formed. There is hardly any sign of smooth \MoTe2 islands.
    From the above observations, it can be concluded that the elevated growth temperature does not support the formation of a smooth \MoTe2 surface. Nevertheless, the obtained structures seem promising and especially the sample with a net of one-dimensional wires encourages further research. The presented Raman spectra in Fig. \ref{fig: tabela_i_schemat}c) indicate that observed wires in Fig. \ref{fig: AFM_serie_pion}h)-i) consist of the \Mo6Te6 \cite{Mo6Te6, Mo6Te6_referee} phase. %This direction requires further research.

    Further MBE growth optimization steps are focused again on the tellurium flux.
    %Another free parameter that was tested in this work is the tellurium molecular flux.
    The panels j) and k) of Fig. \ref{fig: AFM_serie_pion} show AFM scans of samples UW1583 and UW1584. The latter was grown with about three times higher Te flux than it was in the case of the sample UW1583. Apparently, Te flux helped to increase the smoothness of the \MoTe2 layer. Unfortunately, there are still areas of the first monolayer of the \MoTe2 that are not covered, but the second and the third layers are formed.
    The last step was to increase the amount of material to force the complete substrate coverage with \MoTe2  (see Fig. \ref{fig: AFM_serie_pion}l) - UW1585 sample). The coverage is almost 100 \% in this case. The layer is almost continuous. All three samples were grown at 345~\degree C.\\

%%%%Raman

Raman spectroscopy is a technique which helps to determine the phase of the grown \MoTe2 layers \cite{MoTe2_Raman}. In each of the measured samples, very wide lines or a set of lines could be observed. Exemplary Raman spectra of molybdenum telluride grown in relatively high temperature shown in Fig. \ref{fig: tabela_i_schemat}c) with black curve (sample UW1579) suggests the simultaneous occurrence of various phases within a single sample. On the same plot the red color (sample UW1967) corresponds to a rather pure 2H-\MoTe2 phase which forms in lower growth temperatures. Such relation of phase versus growth conditions was indeed observed already in the literature \cite{Raman_phases, MoTe2_Raman_miks_faz}
Another spectroscopy technique, photoluminescence was also used by us to investigate these \MoTe2 samples. In the case of each sample no light emission was observed, which points to dominating nonradiative recombination of excited carriers in our samples.\\

%%%%%%%%%%% XPS %%%%%%%%%%%%%%%%%%%%

\begin{figure}
    \centering
    \includegraphics[width=0.5\textwidth]{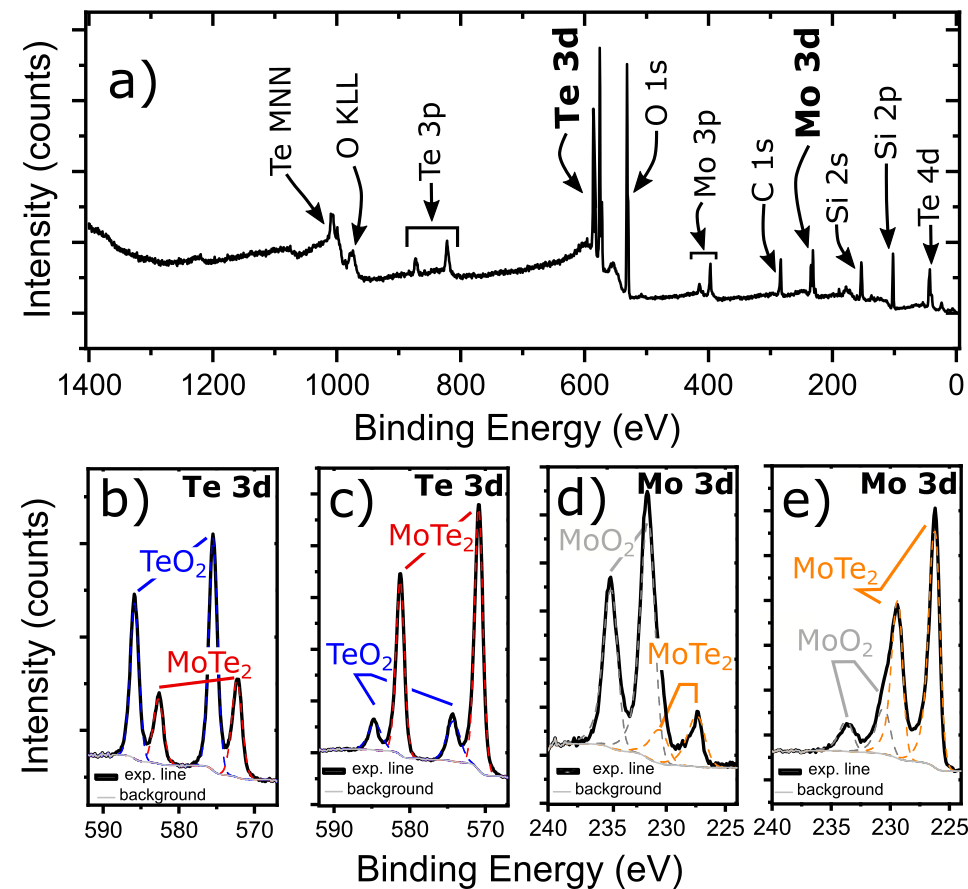}
    \caption{a) XP survey spectrum of UW1577 sample. Characteristic peaks for tellurium, molybdenum, oxygen, silicon and carbon were detected. High resolution spectra in Te 3d and Mo 3d regions reveals presence of TeO$_{\tiny{\textrm{2}}}$ \MoTe2 and MoO$_{\tiny{\textrm{2}}}$  for UW1577 (b,d) and UW1957 (c,e) samples.}
    
    \label{fig: XPS}
 \end{figure}

X-Ray photoelectron spectroscopy (XPS) was utilized to further investigate the surface composition. Survey spectra as well as high resolution spectra in Te 3d and Mo 3d regions were registered for UW1577 and UW1957 samples. Both samples are nominally identical but the latter was grown about 2 years after the UW1577.
Survey scan spectra (Fig. \ref{fig: XPS}a) - only UW1577 sample shown, UW1957 sample similar) confirms that the sample composition was as expected. Only additional small C 1s signal, originating from adventitious carbon (always present in XPS experiments) has been seen.
The high resolution spectra in Te 3d and Mo 3d regions were used to confirm the stoichiometry of deposited \MoTe2 and to distinguish between \MoTe2, MoO$_{\tiny{\textrm{2}}}$ and TeO$_{\tiny{\textrm{2}}}$. Binding energies of each component were taken from the literature \cite{XPS}. 
Additionally using detailed Mo 3d and Te 3d spectra the amount of Mo and Te oxides can be estimated. Based on the registered Mo 3d and Te 3d spectra it can be determined, that in the case of UW1577 sample the oxides’ amounts are much higher than for UW1957 sample. In particular for the UW1577 sample the Te 3d signal assigned to TeO$_{\tiny{\textrm{2}}}$ is much higher than Te 3d signal assigned to \MoTe2 (Fig. \ref{fig: XPS}b) while for the UW1957 sample the opposite can be observed (Fig. \ref{fig: XPS}c). An corresponding relation between Mo 3d signals assigned to oxides and Mo 3d signals assigned to \MoTe2 can be observed for detailed Mo 3d spectra (Fig. \ref{fig: XPS}d) and Fig. \ref{fig: XPS}e). This difference is most probably related to the time between sample preparation and XPS examination. In particular the UW1577 sample has been prepared ca. 2 years before XPS spectra were registered, while the UW1957 sample was freshly grown and XPS analysis was made within a few hours. It suggest that samples oxidizes over time when exposed to air.

%%%%%%%%%%%%%%%%%%%%%%%%%%%%%%%%%%%%%%%%%%%%%%%%%%%%%%
%%%%%%%%%%%%%%%%%%%%%%%%%%%%%%%%%%%%%%%%%%%%%%%%%%%%%%
%%%%%%%%%%%%%%%%%%%%%%%%%%%%%%%%%%%%%%%%%%%%%%%%%%%%%%
% \subsection{Diffusion coefficient discussion} \label{subsect: Diffusion}

    In order to quantitatively compare the influence of the growth conditions on the quality of the samples we use the diffusion coefficient estimation \cite{Mortelmans_difussion}. It relies on the activation energy of adatom diffusion \cite{Diffusion_act_ene} and so, uniquely describes each specific epitaxial system. The relation between the diffusion coefficient $D$ and particular system parameters is the following \cite{Rownanie_dyfuzji} : 
    \begin{equation} \label{eq: diffusion}
    D_{0} e^{(-E_{A}/kT)} = D = 3\theta^{2}/tN^{3}
    \end{equation}
    Here, $E_ {A}$ is the whole \MoTe2 admolecule diffusion activation energy, $N$ is the nucleation density expressed as the number of nucleation seeds per $cm^{2}$, $\theta$ is the surface coverage expressed as the number of adatoms per $cm^{2}$ and $t$ is the growth time. 
    Both $ N $ and $ \theta $ parameters  are easily available from the AFM images. Together with the known growth time $t$ it allows to determine diffusion coefficient $D$ for any system.

In this work, the diffusion constant is discussed for two cases. One is the sample not annealed after the growth - UW1553 shown in Fig. \ref{fig: AFM_serie_pion}c) and the second is the sample, for which the annealing was applied after the growth -  UW1556 shown in Fig. \ref{fig: AFM_serie_pion}d). Other growth parameters are the same. The reason to compare these two samples is to see if the annealing increases the diffusion constant and leads to more smooth sample surface.
The annealing was carried out for 8 hours in the MBE growth chamber, under Te flux. The annealing temperature was about 70 \degree C higher than the growth temperature. 

The surface coverage does not change under annealing. In both cases, it remains at the level of 30\%. Annealing at this temperature does not lead to evaporation of molecules from the substrate surface. Once deposited, the layer of \MoTe2 molecules remains on the surface. However, the number of well-separated islands differs significantly. It decreases from 134 to 79 after annealing in the case of UW1556 sample. It shows that the previously formed islands, as a result of annealing, coalesce to form larger islands, effectively reducing their number, and leading to conclusion that diffusion is enhanced. Quantitative analysis for the sample that was not annealed lead to diffusion coefficient 1.183 $\cdot$ 10$^{-6}$ cm$^{2}$/s. If the same equations are applied to annealed sample, we obtain larger value 5.570 $\cdot$ 10$^{-6}$ cm$^{2}$/s representing effect of two-step process, therefore without direct physical meaning. Still this means that a post-growth annealing enhances the diffusion, as expected. However very different conclusion results from observation of the growth at lower temperatures, which apparently leads to the increase of the diffusion coefficient if Te flux is kept constant. For the sample grown at 50 degrees lower temperature (sample UW1629) the diffusion constant reaches  3.79 $\cdot$ 10$^{-5}$ cm$^{2}$/s. This suggests that decisive factor for diffusion during growth is density of Te atoms on a surface, which is higher for lower substrate temperature for fixed Te flux. 
In each case the value we obtained is at least one order of magnitude higher then discussed in the literature for other TMDC systems \cite{Mortelmans_difussion}. This may be both intrinsic property of the telluride itself or the growth on the atomically smooth hBN.\\

\begin{figure}
    \centering
    \includegraphics[width=0.5\textwidth]{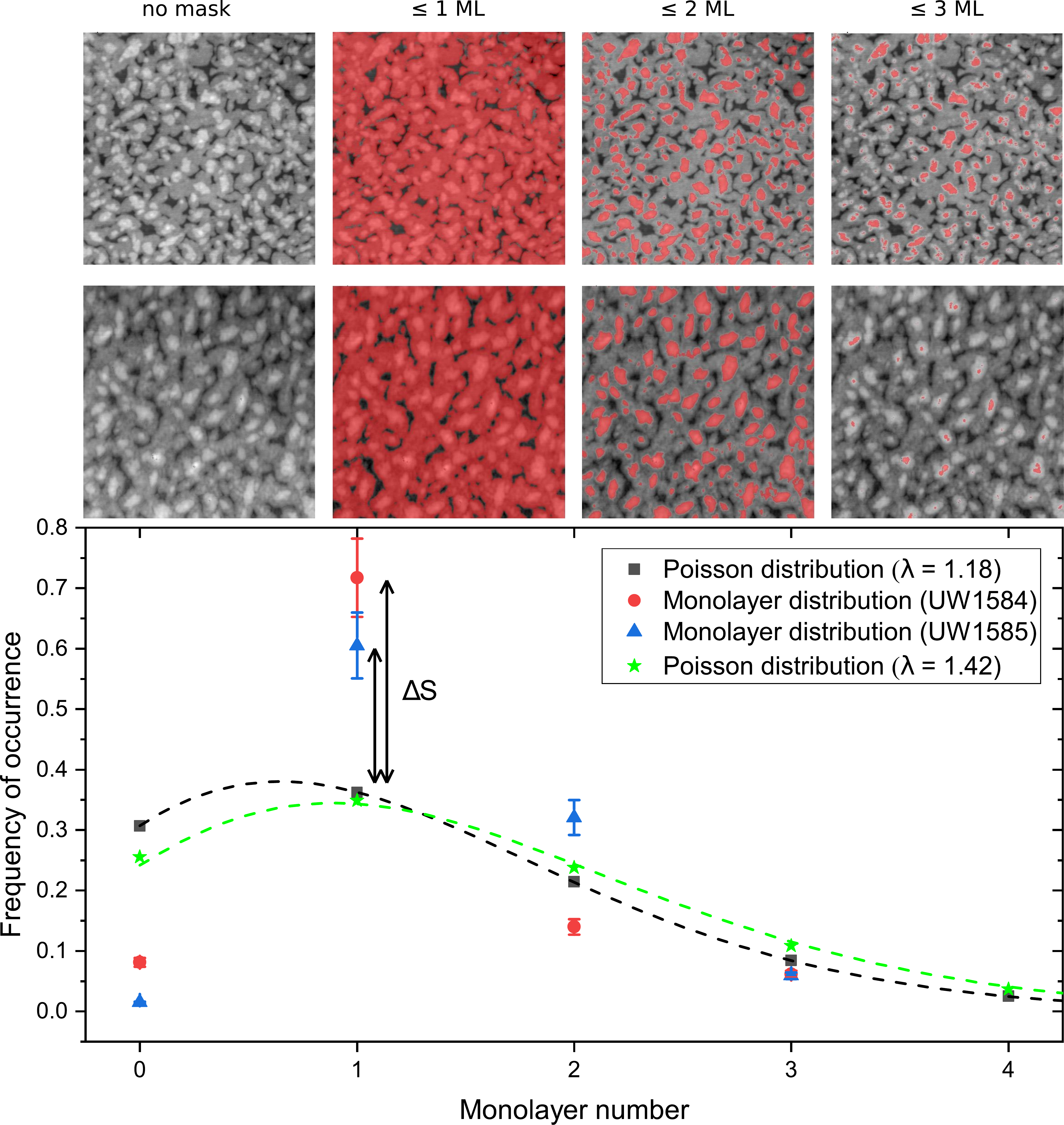}
    \caption{AFM images with masks representing consecutive monolayers of UW1584 and UW1585 samples. Each AFM image is taken for 800 nm $\times$ 800 nm area. Lower panel - the distribution of successive monolayers with the Poisson distribution with the parameter $  \lambda $ equal to the expected value from the distribution. Two distributions for two samples are plotted. $\Delta$S denotes the additional area of the first monolayer obtained by optimizing the growth process.}
    \label{fig: Poisson_UW1584_UW1585}
\end{figure}

To analyze the mobility of admolecules between the successive layers we propose the following model. A single molecule evaporated from the source may fall anywhere on the surface of the substrate. It may be an area completely free of other adatoms, or an area, where an atomic layer has already begun to form.
In particular, the admolecules may end up on the previously formed atomic layer. Then they start to form another layer. Coordinates of the molecule impinging the substrate surface are random. Subsequent molecules settle on the substrate in a certain way, independent of the previous ones. The distributions of these molecules, followed by the distribution of successive atomic layers formed, can therefore be described by a certain probability function. 
Here, we propose to compare such distribution with the Poisson one. The number of $ k $ events in the Poisson distribution equation is the ordinal number of the subsequent monolayer. The $ \lambda $ parameter represents the average number of monolayers for a given MBE growth.
Figure \ref{fig: Poisson_UW1584_UW1585} presents a series of AFM images with masks for samples UW1584 (upper row) and UW1585 (lower row) for subsequent monolayers - 0, 1, 2, 3. The samples differ only in the growth time, 12 h and 15 h respectively. The reason to compare these two is to see if the elongation of the growth leads to the gap filling in the first \MoTe2 layer. Two Poisson distributions are also plotted with the $ \lambda $ parameter equal to 1.12 and 1.26, determining the expected value of the distribution of successive layers for both samples.

In each case the coverage of the first monolayer is greater than predicted by the Poisson distribution. The expected probability of filling 1 ML for both samples, based on the Poisson distribution, is around 0.35. The actual 1 ML coverage for the sample grown for 12 h (UW1584) is 0.72. For the sample grown for 15h (UW1585), the coverage of 1 ML is 0.60.
The substrate areas not covered with material at all are in fact about 3 times smaller than assumed by the Poisson distribution.
The above observations mean that in both cases the growth performed under given conditions gives a more desirable result than random events. The actual coverage of the first monolayer is about twice the expected value. It goes beyond the random event model and tends towards the deterministic growth of a single monolayer of a transition metal dichalcogenide using molecular beam epitaxy. The difference between the expected and actual value of the surface coverage of the first monolayer is marked in Fig. \ref{fig: Poisson_UW1584_UW1585} as $\Delta$S.

Comparing the distribution of successive monolayers of the UW1584 and UW1585 sample, it can be seen that a longer growth does not result in a greater coverage of the first monolayer. Rather, the prolongation of the growth time causes the second monolayer to grow at the expense of the first one.

\section{Conclusions}

The MBE growth of \MoTe2 on hBN leads to different structures depending on the growth conditions. It is possible to obtain a network of quasi one-dimensional objects, as well as a smooth layer with a coverage similar to a single monolayer.
The diffusion coefficient allows to quantify the effect of the growth conditions on the dynamics of molecules on the surface.
Thanks to the analysis of the distribution of the population of subsequent \MoTe2 layers on hBN, it is evident that this growth is far from accidental represented by Poisson distribution. By manipulating the growth parameters, it is possible to exit the regime of accidental deposition of the material on the substrate in favor of formation of the single monolayer.

The formation of large, continuous, monocrystalline phase controlled \MoTe2 layers still remains a challenge. The search for optimal growth conditions and suitable substrates still leaves a lot of room for further investigations. 

\section{Acknowledgements}
We gratefully acknowledge the financial support from projects no. 2017/27/B/ST5/02284, 2020/39/B/ST7/03502, 2021/41/B/ST3/04183 and 2019/33/B/ST5/02766 financed by the National Science Centre (Poland).

%\bibliography{bibliography}

\end{document}